# Experimental realization of chiral Landau levels in two-dimensional Dirac cone systems with inhomogeneous effective mass

Hongwei Jia[1,2,#,*], Mudi Wang[1,#], Shaojie Ma[3], Ruo-Yang Zhang[1], Jing Hu[1], C. T. Chan[1,†]

[1]Department of Physics, the Hong Kong University of Science and Technology, Clear Water Bay, Kowloon, Hong Kong, China

[2]Institute for Advanced Study, the Hong Kong University of Science and Technology, Clear Water Bay, Kowloon, Hong Kong, China

[3]Department of Physics, University of Hong Kong; Hong Kong, China

[#]these authors contributed equally to this work

[*]jiahongwei7133@gmail.com, [†]phchan@ust.hk

Abstract: Chiral zeroth Landau levels are topologically protected bulk states that give rise to chiral anomaly. Previous discussions on such chiral Landau levels are based on three-dimensional Weyl degeneracies. Their realizations using two-dimensional Dirac point systems, being more promising for future applications, were never reported before. Here we propose a theoretical and experimental scheme for realizing chiral Landau levels in a photonic system. By introducing an inhomogeneous effective mass through breaking local parity inversion symmetries, the zeroth-order chiral Landau levels with one-way propagation characteristics are experimentally observed. In addition, the robust transport of the chiral zeroth mode against defects in the system is experimentally tested. Our system provides a new pathway for the realization of chiral Landau levels in two-dimensional Dirac systems, and may potentially be applied in device designs utilizing the transport robustness.

Linear degeneracies in band structures are singularities in momentum space. These singularities are the sources of topological characteristics in band theory. Typical examples are the Dirac and Weyl points [1-10,13], which are linear crossings of two bands. They give rise to nontrivial topological invariants, which are associated with a variety of novel phenomena, such as topologically protected gapless surface states that connect two singular points with opposite chiralities [6,8,9,10,13]. Singular points with opposite topological charges can annihilate each other [1,11]. Since the dynamical equation describing the Weyl or Dirac cones has a form similar to that of relativistic Fermions in quantum field theory [12], such systems are good platforms for simulating relativistic particles and observing the single particle behaviors. Recent advances of inhomogeneous modulations on unit cell morphologies successfully introduced synthetic gauge fields to the quasi-particles [14-24]. As a result, the eigen-energies of bulk states become quantized, leading to the Landau levels. Associated phenomena such as flat bands and chiral Landau levels have been experimentally detected in photonic and phononic systems [15,16,20,22-25].

Different from topological surface states, the chiral Landau level is a one-way propagative bulk state [12,20,25,26], which is also topologically protected. In quantum field theory and condensed matter physics, chiral Landau level plays an important role in chiral anomaly [26-28], characterized by the non-conservation of chiral currents. It is commonly believed that the chiral Landau levels arise from the interaction between 3D Weyl degeneracies and background magnetic fields. Compared with 3D Weyl systems, 2D Dirac systems are more accessible for fabrication, and thus are more promising for future applications. However, realizing chiral Landau levels with 2D Dirac degeneracies has not yet been considered.

In this work, we propose the realization of chiral Landau levels using a photonic honeycomb system. By breaking local parity inversion symmetry in each unit cell, an inhomogeneous effective mass is introduced into the Dirac cone, which is equivalent to an in-plane gauge field coupled with the Dirac quasi-particles. As a result, the energy levels become quantized, and in-plane chiral Landau levels arise, which are one-way propagative and is robust against perturbations in the bulk. With the inhomogeneous platform, the band dispersions of Landau levels are experimentally measured. In addition, the robustness of transport of the zeroth chiral mode, is also experimentally tested by introducing disordered defects to the system.

In a honeycomb lattice system with both parity inversion and time reversal symmetries (see Fig. 1a), the energy bands cross linearly at the $K$ and $K'$ points in the Brillouin zone (see Fig. 1b). Here, the lattice sites $A$ and $B$ are cylinders with a high dielectric constant ($\varepsilon_m=13.8$), and the diameters are denoted by $d_A$ and $d_B$, respectively. The other parts are filled with air ($\varepsilon_{air}=1$). The cone-like dispersions near the $K$ or $K'$ points can be described by the 2D Dirac Hamiltonian $H = v(k_x\sigma_x \pm k_y\sigma_y)$, where $v$ is the group velocity near the Dirac point, $\sigma_{x,y,z}$ are Pauli matrices, and $k_x$ and $k_y$ are Block wave vectors relative to $K$ or $K'$ in $x$ and $y$ directions, respectively. A typical way to introduce synthetic gauge fields is to tune the position of the degeneracy points in **k**-space. However, the artificial magnetic field generated in this way is perpendicular to the $xy$ plane, which results in flat-band Landau levels for such 2D systems [15,16,22-24]. To obtain an in-plane magnetic field and consequently inducing in-plane dispersive Landau levels, we need to go beyond the conventional methods. Here, we locally break parity inversion symmetry by setting $d_A \neq d_B$ to introduce an effective mass $m$. This mass term will lift the Dirac degeneracy and open a bandgap locally (the gap size $\Delta\omega = 2|m|$). If the gap size varies with position, one obtains an effective Hamiltonian that describes the interaction between the Dirac quasi-particles and the synthetic gauge field

$$H' = v(\hat{k}_x\sigma_x \pm \hat{k}_y\sigma_y) + m(\mathbf{r})\sigma_z \ . \tag{1}$$

Such a Hamiltonian near the $K/K'$ points can be realized by the hexagonal structure illustrated in Fig. 1(c), in which $d_A$ varies from 2 to 6mm from the left to the right of the sample ($x$ direction), and conversely $d_B$ varies from 6 to 2mm. Periodic boundary condition is applied in the $y$ direction. The sign $\pm$ in Eq. (1) correspond to $K$ and $K'$ valleys, respectively. $\hat{k}_{x,y}$ with hats denote wave vector operators, and in the direction without translational symmetry $\hat{k}_x = -i\partial_x$. In the other direction, translational symmetry is preserved, and thus $\hat{k}_y = k_y$ is still a good quantum number. The variation step between the same lattice sites is fixed at $\Delta d_A$=0.1mm (and $\Delta d_B$=−0.1mm), and thus $d_A$=$d_B$=4mm on the line $x$=0 of the sample. This graded morphology causes the gap size $\Delta \omega$ to be linearly dependent (approximately) on the coordinate $x$ (as shown in Fig. 1c), which is equivalent to adding an effective mass term to the Dirac Hamiltonian that is linear with respect to $x$ ($m = ax$). The photo of part of the fabricated sample for experimental measurement is displayed in Fig. 1(d). Since the effective mass introduces a $\sigma_z$ term in the Hamiltonian [Eq. (1)], it is equivalent to a vector potential in $z$ direction [$A_z$=$m(x)$], meaning that an effective canonical momentum $\hat{k}_z = k_z + A_z$ (with $k_z$=0) is introduced into the system. Therefore, we expect to see phenomenon that arise from a synthetic magnetic field in the $y$ direction (as indicated by the yellow arrow in Fig. 1d), which is determined by the definition of the magnetic field $B_y = [\hat{k}_x, \hat{k}_z]$.

The designed sample preserves translational symmetry in the $y$ direction, and the synthetic magnetic field is also in $y$ direction. The synthetic gauge field will result in the quantization of energy levels, which are expressed as (see Supplemental Materials [29] for derivation details)

$$\omega_n = \begin{cases} \chi v k_y, & n = 0 \\ \pm\sqrt{v^2 k_y^2 + 2n|a|v}, & n \geq 1 \end{cases} \quad (2)$$

Here $\chi = \pm\text{sgn}(a)$ with $a$>0 for our sample, and $\pm$ correspond to $K$ and $K'$ valleys, respectively. From this equation, we see that the energy levels are dispersive in $k_y$ direction, and the zeroth order Landau level has a linear dispersion and can have group velocities in +$y$ (up-going) or –$y$ (down-going) directions, which is determined by $K$ or $K'$ valleys, i.e. the chirality of the Dirac point. For the experimental sample, the band structure of the supercell under periodic boundary conditions in $y$ direction (Fig. 1c) can be interpreted as the Landau levels under the action of a synthetic gauge field, and the full wave simulation result (near $k_y$=0, the blue dot) is plotted in Fig. 1e. Here the dotted lines show high order Landau levels, and the linearly dispersive zeroth modes are labelled by blue and red lines, which are affiliated to $K$ and $K'$ valleys, respectively. The analytic expressions of Landau levels (Eq. 2) can well reproduce the band structure if the $k_y$ domain is the vicinity of the $K$ or $K'$ points (see Section 2 of Supplemental Materials [29] for details). We note that the boundary condition in the $x$ direction does not affect the computed dispersions, because the field of the eigenstates in the frequency range of interest are confined near $x=0$ (see later experimental results). To experimentally measure the Landau levels, we put a point source at the bottom center of the sample [see the setup in the left panel of Fig. 2(a)], which excites the up-going propagating modes in the bulk [Fig. 2(a), right panel]. With a discrete Fourier transformation, the energy level dispersions can be obtained experimentally, and the Fourier transformed field intensity $|E_z|$ as a as a function of $k_y$ and $f$ is shown in Fig. 2b. The result is consistent with full wave simulations and theoretical predictions (see Supplemental Materials [29] for details). We see that for the zeroth order Landau level, only the positively propagating mode is excited for the excitation condition in Fig, 2(a). The dispersion is linear, as expected. By contrast, if we locate the source at the top center [see the setup in the left panel of Fig. 2(c)], the zeroth mode that is propagating in $-y$ direction can be excited (right panel of Fig. 2c). Via a discrete Fourier transformation, the experimental result of the energy dispersions can be resolved, as shown in Fig. 2(d). The field distribution $|E_z|$ of the zeroth order chiral mode in $x$ direction is also measured, as shown in Fig. 2(e). It is seen that the mode amplitude is localized near the middle ($x=0$) of the sample. Since the up-going and down-going zeroth chiral mode is affiliated to $K$ and $K'$ points respectively, the electromagnetic response of $K$ and $K'$ points are different from each other if the excitation frequency is within the bandgap between ±1st order Landau levels. Relevant discussions are shown in Section 2 of Supplemental Materials [29]. The observed phenomenon comply with the symmetry of the system. Even though the local parity inversion symmetry of each unit cell is broken, the overall sample still has a parity inversion symmetry. The sources at the bottom center and top center can be transformed to each other via parity inversion operations. Correspondingly, and the positively and negatively propagating chiral modes excited by the two sources (as well as $K$ and $K'$ points) are also parity-symmetric to each other.

The chiral zeroth Landau level is protected against small perturbations. Next, we provide a more detailed experimental test for the robustness of transport against imperfections. We first put a point source at the center of the bottom edge of the sample, and at the same time, some defects (4 additional cylinders with a diameter of 3mm) are introduced at the center of the sample, as shown in

Fig. 3(a). The zoomed in picture of the defects is shown in the inset. Compared with Fig. 2(e), we find that the defects almost span the entire confined region of the field of the zeroth mode, and such defects will result in strong reflections for ordinary propagating modes. However, even though localization of the field strength distribution ($|E_z|$) can be observed at the center of the sample (location of the defect) from the simulation result in Fig. 3(b), the chiral zeroth mode bypasses the defects and the reflected field intensity is negligibly small compared with the transmitted field intensity. We next measure dispersion of the Landau levels to quantitatively test the reflection strength. By locating the source at the bottom center, and the field distribution ($|E_z|$) in $k_y$-$f$ space can be resolved experimentally. It is shown that the zeroth mode that is propagating in $-y$ direction is almost not excited compared with that in $+y$ direction even though the defect is introduced, as indicated by the dispersions in Fig. 3(c). The experimental results provide solid evidence for the robustness of the transport of the chiral zeroth Landau level. The weak backscattering of the chiral zeroth Landau level can be intuitively understood. Since the zeroth modes with positive and negative group velocities are affiliated to different valleys (i.e. *K* and *K'*), the backscattering of the zeroth mode is essentially an inter-valley scattering if the illumination frequency is inside the gap. Since the *K* and *K'* points are widely separated in **k**-space, the inter-valley coupling is weak.

To summarize, we propose a theoretical and experimental scheme for the realization of chiral zeroth Landau levels via inhomogeneously breaking local parity inversion symmetry in 2D Dirac point systems. Such a system is experimentally realized using a photonic honeycomb lattice system, in which the diameter of lattice sites in each unit cell are tuned to introduce an effective mass that depends linearly on the spatial coordinate. Based on such a platform, the in-plane Landau level dispersions are experimentally measured. In addition, the robustness of the zeroth order mode is also experimentally demonstrated by introducing defects at the center of the sample. Our proposal extends the chiral zero Landau level from 3D systems to 2D systems, which are much easier to fabricate, and are hence more promising for applications. By scaling the sample from millimeter region to nanometer region, the scheme may inspire the design of low-loss photonic crystal components that can utilize the robustness of the chiral Landau levels.

Acknowledgements: This work is supported by Research Grants Council of Hong Kong through grants AoE/P-502/20, 16307621, 16307821, 16307420, 16310420 and Croucher Foundation (CAS20SC01) and KAUST20SC01. Y. Zhu acknowledges the financial support from National Natural Science Foundation of China (NSFC) grant 11701263.

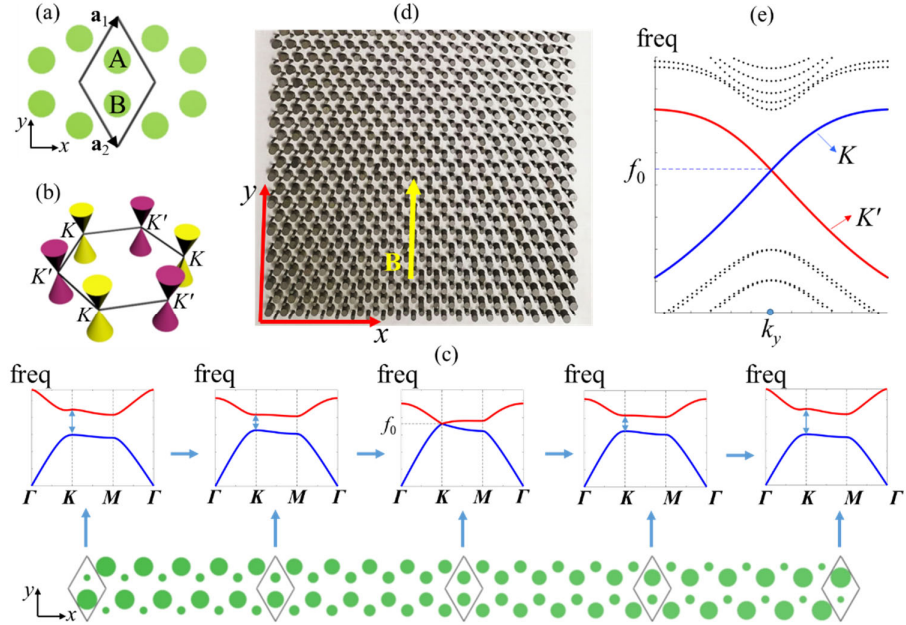

Fig. 1. Design of the honeycomb lattice carrying synthetic gauge fields. (a) Structure of the honeycomb lattice, with A and B being lattice sites made of cylinders with a high dielectric constant ($\varepsilon_m$=13.8). The unit cell is indicated by the black rhombus, with $\mathbf{a}_1$ and $\mathbf{a}_2$ being the lattice vectors. (b) Brillouin zone of the honeycomb lattice, and the band dispersions form Dirac cones near $K$ or $K'$ pints. (c) Design of the supercell of the sample. The sample is non-periodic in $x$ direction and periodic in $y$ direction. For each unit cell, local parity inversion symmetry is broken by tuning the difference of diameter $\Delta d = d_A - d_B$ between cylinders $A$ and $B$. $\Delta d$ varies linearly from positive to negative from left to right. As a consequence, the bandgap $\Delta f$ at $K$ (and $K'$) varies linearly (approximately) as indicated by the band structures corresponding to the local structure. (d) A photograph of part of the sample, the inhomogeneous effective mass is equivalent to an artificial magnetic field $\mathbf{B}$ in $y$-direction (labeled by the yellow arrow). (e) Full wave simulation of band structure of the supercell structure under periodic boundary condition in $y$-direction, with dotted lines representing higher order modes, and the blue and red lines are zeroth modes affiliated to $K$ and $K'$, respectively.

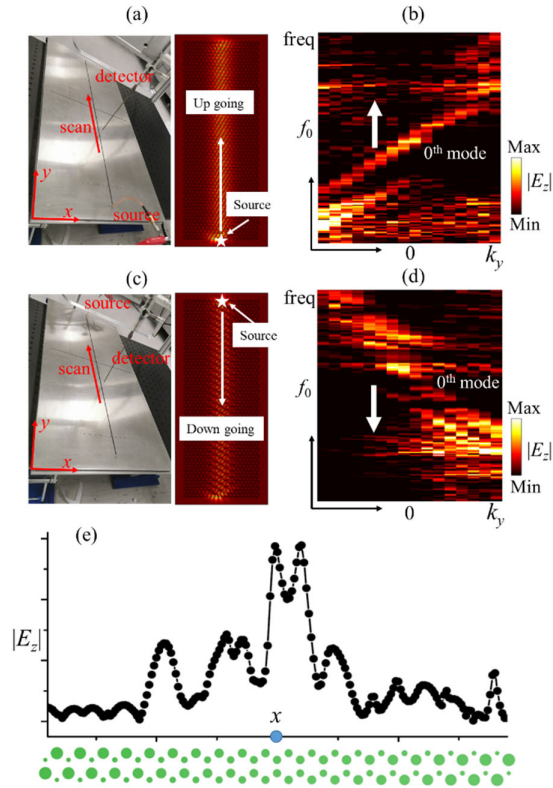

Fig. 2. Experimental measurement of chiral Landau levels. (a) Left panel: Putting the point source with an out of plane polarization at the bottom center of the sample to excite the up-going (+$y$ direction) chiral Landau level. Right panel: Field strength $|E_z|$ distribution (simulation result) corresponding to the excitation in the left panel (b) Dispersion of Landau levels obtained by experimental measurement under the excitation condition in (a). The up-going chiral zeroth mode is excited. (c) Left panel: Locating the point source with an out of plane polarization at the bottom center of the sample to excite the down-going (−$y$ direction) chiral Landau level. Right panel: Field strength $|E_z|$ distribution (simulation result) corresponding to the excitation in the left panel (d) Dispersion of Landau levels obtained by experimental measurement under the excitation condition in (c). (e) Field strength distribution along $x$-direction ($|E_z|$) of zeroth order chiral Landau level obtained from experimental measurement.

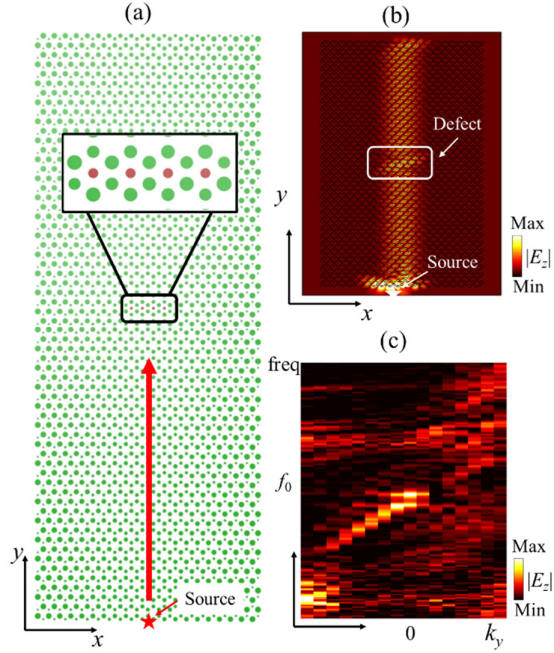

Fig. 3. Experimental test on the transport robustness of the chiral zeroth Landau level. (a) Putting some defects at the center of the sample, which spans almost the domain of the field distribution ($x$ direction) of the zeroth mode to disrupt its propagation. A point source is placed at the bottom center of the sample. (b) Full wave simulation of magnitude of the field ($|E_z|$) under the excitation condition in (a). Field localization can be observed within the defect domain, but the reflection is almost negligible. (c) Experimental result of Landau level dispersions resolved from the field distributions with defect in (a). The up-going zeroth mode is dominantly excited, and the excitation of down-going mode due to reflection at the defect is negligible.

# Supplemental Material: Experimental realization of chiral Landau levels in two-dimensional Dirac cone systems with inhomogeneous effective mass

Hongwei Jia[#,*], Mudi Wang[#], Shaojie Ma, Ruo-Yang Zhang, Jing Hu, C. T. Chan[†]

## 1. Derivation of Landau level dispersions

Here we derive the dispersion relation of Eq. (2) in the main text. We start from the Dirac equation coupled with the inhomogeneous effective mass (Eq. 1 in the main text), in which the effective mass is linear with respect to $x$-direction.

$$H' = v(\hat{k}_x \sigma_x \pm \hat{k}_y \sigma_y) + ax\sigma_z \tag{S1}$$

As the Hamiltonian includes Pauli matrices, we firstly need to square the Hamiltonian

$$[v(\hat{k}_x \sigma_x \pm \hat{k}_y \sigma_y) + ax\sigma_z]^2 \varphi = \omega^2 \varphi$$
$$\Rightarrow [v^2 \hat{k}_x^2 + v^2 \hat{k}_y^2 + (ax)^2 - iav\sigma_y[\hat{k}_x, x]]\varphi = \omega^2 \varphi \tag{S2}$$

Here we used the following commutation relations

$$[\hat{k}_x, \hat{k}_y] = 0, \ [\hat{k}_y, x] = 0 \tag{S3}$$

Since $\hat{k}_x = -i\partial_x$, it is not difficult to find the commutation

$$[\hat{k}_x, x] = -i\partial_x x = -i \tag{S4}$$

Based on the commutation relation, we can define the creation and annihilation operators

$$b^\dagger = N(v\hat{k}_x - i|a|x), \ b = N(v\hat{k}_x + i|a|x) \tag{S5}$$

The commutation relation $[b^\dagger, b] = 1$ requires that

$$1 = [N(v\hat{k}_x - i|a|x), N(v\hat{k}_x + i|a|x)] = N^2([v\hat{k}_x, i|a|x] + [-i|a|x, v\hat{k}_x])$$
$$= 2iv|a|N^2[\hat{k}_x, x] = 2N^2 v|a| \tag{S6}$$
$$\Rightarrow N = \frac{1}{\sqrt{2v|a|}}$$

Next we define the particle number operator

$$\hat{n} = b^\dagger b = N^2(v\hat{k}_x - i|a|x)(v\hat{k}_x + i|a|x)$$
$$= N^2(v^2\hat{k}_x^2 + a^2x^2 + [v\hat{k}_x, i|a|x]) = N^2(v^2\hat{k}_x^2 + a^2x^2 + v|a|) \tag{S7}$$

and we thus have

$$\frac{1}{N^2}\hat{n} = v^2\hat{k}_x^2 + a^2x^2 + |a|v \tag{S8}$$

Eq. (S2) reduces to

$$[\frac{1}{N^2}\hat{n} - |a|v - av\sigma_y + v^2\hat{k}_y^2]\varphi = \omega^2\varphi \tag{S9}$$

There is still a $\sigma_y$ term in Eq. (S9), and we need the following process

$$\sqrt{[\frac{1}{N^2}\hat{n} - |a|v - \omega^2 + v^2\hat{k}_y^2]^2}\varphi = \sqrt{(av\sigma_y)^2}\varphi \tag{S10}$$
$$\Rightarrow [2|a|v\hat{n} - |a|v - \omega^2 + v^2\hat{k}_y^2]\varphi = \pm|a|v\varphi$$

It is then not difficult to derive the analytical expression of Landau levels

$$\omega_n = \begin{cases} \pm\sqrt{v^2 k_y^2 + 2n|a|v} & n \geq 1 \\ \chi v k_y & n = 0 \end{cases} \tag{S11}$$

The Landau level dispersions in Eq. (2) of the maintext are thus obtained.

## 2. Numerical simulations

In calculating the Landau levels of the Dirac points with inhomogeneous effective mass, the $k_y$ component of the Block wave vector should be in the vicinity of the $K$ or $K'$ points. This means that Eq. (S11) is valid only if $k_y$ is near the $\Gamma$ points (i.e. $k_y$ is near $2n\pi/\sqrt{3}a$, $n \in \mathbb{Z}$) in the **k**-space of the supercell system. However, the complete band structure of the supercell under periodic boundary condition in $y$ direction can be numerically obtained. In Fig. S1(a) (also Fig. 1e in the main text), we show the simulation results of the band structure calculated by the wave optics module of COMSOL. Landau level dispersions can be easily identified from the figure. It is also found that zeroth modes with both positive and negative group velocities are in the band structure, because the results includes the Landau levels of all the valleys (both $K$ and $K'$ points). Each of them is affiliated to a $K$ or a $K'$ point in the Brillouin zone of the 2D honeycomb lattice (the inset of Fig. S1a), as specified in the figure. For comparison, we also show the theoretical results of Landau level dispersions obtained by Eq. S11, as displayed in Fig. S1(b-c) for $K$ and $K'$, respectively. As expected, the analytical expression can well reproduce profile of the band dispersion if $k_y$ is near $2n\pi/\sqrt{3}a$, but derivate from the full wave simulation results if $k_y$ is away from $2n\pi/\sqrt{3}a$. This is because the Dirac Hamiltonian can only describe the cone-like dispersions near $K$ or $K'$ points.

In the main text, the Landau level dispersions are experimentally measured under different excitation conditions. Within the bandgap between ±1st order Landau levels, the electromagnetic response of $K$ and $K'$ valley are different from each other, owing to the fact that the zeroth mode is either up-going or down-going and affiliated to different valleys (see Fig. S1d). Therefore, if the source is located at the bottom center to excite the up-going zeroth mode (see Fig. 2a-b), only the eigenstates at $K$ point can be excited. As indicated in Fig. S1e, the field strength $|E_z|$ at $K$ points is much higher than that at $K'$ at any frequency inside the gap (here we set the frequency to the central frequency). Conversely, if the source is located at the top center to excite the down-going chiral zeroth mode, only the eigenstates at $K'$ points can be excited, and resultantly, the field strength $|E_z|$ at $K'$ point is much higher than that at $K$ point (see Fig. S1f). The results in Fig. 1Se-f are obtained numerically.

The honeycomb lattice is a typical 2D system, with its structure homogeneous in $z$ direction and extends to infinity. However, such a structure cannot be realized in experiment, and thus mostly we truncate the structure and apply perfectly electric conductor (PEC) boundary condition in $z$ direction, so that the structure can have a finite size. Such a truncation of the sample does not change profile of the band dispersions because the symmetries of the structure remain unchanged. However, the frequency will be shifted a little. In our system, the central frequency obtained numerically is $f_0$=7.3GHz, which is shifted to $f_0$=8GHz in experiment.

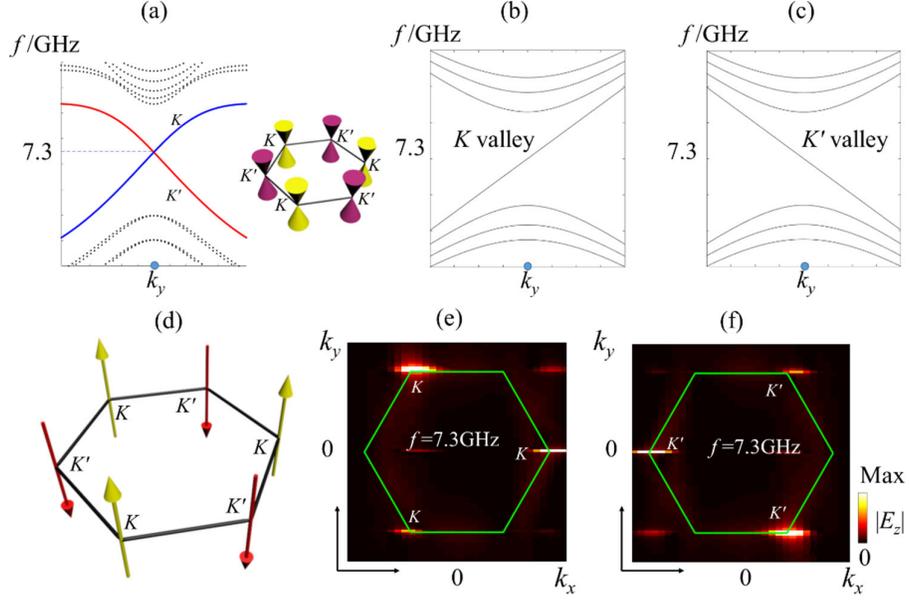

Fig. S1. Comparison of Landau levels between full wave simulation and analytical expressions. (a) Full wave simulation of the band structure of the supercell system in Fig. 1(c) under periodic boundary condition in $y$ direction. Results are obtained with the wave optics module of COMSOL. The band structure can be interpreted as the Landau levels, with the zeroth modes having a positive group velocity affiliated to $K$ valley, and those having a negative group velocity affiliated to $K'$ valleys (as specified in the figure). (b-c) Landau level dispersions obtained by Eq. (S11). The results can well predict the band structures if $k_y$ is near the $K$ or $K'$ points (i.e. $k_y$ is near $2n\pi/\sqrt{3}a$, $n \in \mathbb{Z}$, labelled by blue dots), and diverges from the simulation result if $k_y$ is away from $2n\pi/\sqrt{3}a$.